\documentclass[prb,superscriptaddress,reprint,showpacs,amssymb,amsmath,floatfix,aps]{revtex4-1}

\usepackage{graphicx}                             

\usepackage{color}
\definecolor{red}{rgb}{1.00,0.00,0.00}
\usepackage{multirow}
\usepackage{soul}

\date{\today}

\begin{document}

\title{Strain in epitaxial MnSi films on Si(111) in the thick film limit studied by\\
polarization-dependent extended x-ray absorption fine structure}

\author{A. I. Figueroa}
\affiliation{Magnetic Spectroscopy Group, Diamond Light Source, Didcot, OX11~0DE, United Kingdom}

\author{S. L. Zhang}
\affiliation{Clarendon Laboratory, Department of Physics, University of Oxford, Parks Road, Oxford, OX1~3PU, UK}

\author{A. A. Baker}
\affiliation{Magnetic Spectroscopy Group, Diamond Light Source, Didcot, OX11~0DE, United Kingdom}
\affiliation{Clarendon Laboratory, Department of Physics, University of Oxford, Parks Road, Oxford, OX1~3PU, UK}

\author{R. Chalasani}
\affiliation{Department of Materials Science and Engineering, Tel Aviv University, Ramat Aviv 6997801, Tel Aviv, Israel}

\author{A. Kohn}
\affiliation{Department of Materials Science and Engineering, Tel Aviv University, Ramat Aviv 6997801, Tel Aviv, Israel}

\author{S. C. Speller}
\affiliation{Department of Materials, University of Oxford, Parks Road, Oxford, OX1~3PH, UK}
 
\author{D. Gianolio}
\affiliation{Diamond Light Source, Didcot, OX11~0DE, United Kingdom}

\author{C. Pfleiderer}
\affiliation{Physik-Department, Technische Universit\"{a}t M\"{u}nchen, D-85748 Garching, Germany}

\author{G. \surname{van der Laan}}
\affiliation{Magnetic Spectroscopy Group, Diamond Light Source, Didcot, OX11~0DE, United Kingdom}

\author{T. Hesjedal}
\email[Corresponding author: ]{Thorsten.Hesjedal@physics.ox.ac.uk}
\affiliation{Clarendon Laboratory, Department of Physics, University of Oxford, Parks Road, Oxford, OX1~3PU, UK}

\date{\today}

\begin{abstract}
We report a study of the strain state of epitaxial MnSi films on Si(111) substrates in the thick film limit (100-500~\AA) as a function of film thickness using polarization-dependent extended x-ray absorption fine structure (EXAFS). All films investigated are phase-pure and of high quality with a sharp interface between MnSi and Si. The investigated MnSi films are in a thickness regime where the magnetic transition temperature $T_\mathrm{c}$ assumes a thickness-independent enhanced value of $\geq$43~K as compared with that of bulk MnSi, where $T_\mathrm{c} \approx 29~{\rm K}$.
A detailed refinement of the EXAFS data reveals that the Mn positions are unchanged, whereas the Si positions vary along the out-of-plane [111]-direction, alternating in orientation from unit cell to unit cell.
Thus, for thick MnSi films, the unit cell volume is essentially that of bulk MnSi --- except in the vicinity of the interface with the Si substrate (thin film limit). 
In view of the enhanced magnetic transition temperature we conclude that the mere presence of the interface, and its specific characteristics, strongly affects the magnetic properties of the entire MnSi film, even far from the interface. Our analysis provides invaluable information about the local strain at the MnSi/Si(111) interface. The presented methodology of polarization dependent EXAFS can also be employed to investigate the local structure of other interesting interfaces.
\end{abstract}


\maketitle

\section{\label{Introduction}Introduction}

Skyrmions are topologically non-trivial whirls of the magnetization. They have been observed as individual objects as well as in the form of periodic, three-dimensional lattices with periodicities in the range $\sim$$2 -100\,{\rm nm}$, covering nearly two orders of magnitude.\cite{Muhlbauer09,Pfleiderer10,Nagaosa13} 
Magnetic skyrmions were first identified in bulk compounds with non-centrosymmetric crystal structure, such as the B20 family of crystals, in which the Dzyaloshinskii-Moriya spin-orbit interactions play a crucial role.  Using small angle neutron scattering,\cite{Muhlbauer09,Adams2011} and magnetotransport measurements (topological Hall effect),\cite{Neubauer09} the non-trivial topological character of the skyrmion lattice could be confirmed. However, it was in particular the observation that skyrmions in bulk samples can be easily manipulated with current densities that are nearly six orders of magnitude smaller than those needed for conventional spin transfer torque (STT) based schemes (currently $\sim$10$^{11}$~A m$^{-2}$) that has generated great interest in skyrmions as a novel route towards applications.\cite{Jonietz10} 

First studies in thinned bulk samples using real space imaging by means of Lorentz transmission electron microscopy (LTEM) early on suggested that skyrmions tend to be favored in systems of reduced size.\cite{Yu10,Yu11,Tonomura12} In addition, skyrmions have also been detected by means of spin-resolved scanning tunneling microscopy in thin films grown on carefully selected substrates mediating strong antisymmetric spin interactions.\cite{Heinze11} 
 
For future applications such as skyrmionic memory and logic devices,\cite{VA} the development of skyrmion-carrying thin films and nanomaterials on common semiconductor substrates is a crucial prerequisite.\cite{Tokura_review_skyrmion_Natnano_13}
As the non-centrosymmetric B20 phase of MnSi is among the most extensively studied skyrmion systems, thin films on a Si substrate appear to be an obvious choice --- at least for studies of fundamental aspects.\cite{Muhlbauer09,Jonietz10,Milde13,Nagaosa13,Schwarze2015} In fact, the skyrmion lattice phase is not only observed in MnSi bulk crystals, but also in many other Si- and Ge-based B20 systems that are helimagnetically ordered in zero magnetic field, such as Fe$_{1-x}$Co$_x$Si,\cite{Muenzer2010} Mn$_{1-x}$Fe$_x$Si, Mn$_{1-x}$Co$_x$Si,\cite{Pfleiderer10} and related Ge-based compounds. 

Phase-pure MnSi films on Si(111) have been successfully grown by molecular beam epitaxy (MBE).\cite{Karhu10,Karhu11,Karhu12,Wilson12,Wilson13,Li13,Menzel13,Wilson14,Yokouchi15,Zhang_MnSi_2016} Yet, despite of the high quality of the films reported in the literature, it is still unresolved if a skyrmion lattice phase forms in MnSi epilayers.\cite{Meynell2014} A first important difference between thin bulk samples and epilayers concerns the helimagnetic transition temperature, which is reduced from 30\,K to 25\,K in the former, whereas increased values as high as $>$80~K are consistently observed for thin films on Si(111).\cite{Zhang_MnSi_2016}

Further, for the applied magnetic field perpendicular to the epitaxial films, early polarized neutron reflectometry and magnetization measurements, eventually led to the conclusion that skyrmions do not form. This was challenged by LTEM studies and Hall effect measurements,\cite{Li13} which initially appeared to provide putative evidence for skyrmions, but eventually seemed to confirm the absence of skyrmions.\cite{Meynell2014} In contrast, for the magnetic field parallel to the film, it has been suggested that skyrmion cigars form as conjectured from magnetization measurements as well as recent studies of the planar Hall effect. 

In view that individual skyrmions as well as skyrmion-lattice order have been observed in thinned bulk samples with a thickness comparable to MBE-grown epilayers, the interface with the Si substrate must be an important source for the differences of the magnetic properties of bulk samples and epilayers. 

It is conceivable that the interface may thereby influence the properties of the thin films in at least two prominent ways.
First, it has recently been suggested theoretically, that the mere existence of a surface already generates additional terms in the free energy that modify the magnetic order within a considerable distance to the interface.\cite{Mueller2016} Second, the lattice mismatch generates strain and it is the response to this strain that drives modifications of the magnetic properties.\cite{Wilson14}

Two variants of lattice strain may be distinguished, notably isotropic strain as generated by hydrostatic pressures and anisotropic strain generated by uniaxial compressive or tensile pressures. The former has long been explored in great detail in bulk samples, where a suppression of the helimagnetic transition temperature is observed, and more recently in thin films.\cite{Engelke2014}
In contrast, the effects of uniaxial pressure have only been investigated recently in detail for bulk MnSi using small-angle neutron scattering.\cite{Chacon2015,Nii2015} 

In order to resolve the controversies surrounding the magnetic phases in epitaxial MnSi thin films on Si(111) a first and important step are detailed measurements of the in-plane and out-of plane lattice parameters of the films. In particular, this represents a precondition for strain engineering of thin films with higher transition temperatures and tailored skyrmion lattices.\cite{Zhang_MnSi_2016}

For our study we used extended x-ray absorption fine structure (EXAFS) to determine the local atomic environment of Mn atoms in MnSi thin films on Si(111) in order to infer the presence and size of local strain. EXAFS is a powerful tool for quantifying the variation of interatomic distances, which has been successfully used to study surface and interfaces structure in, e.g., MnSi monolayers.\cite{Kahwaji12} The polarization dependence of the EXAFS cross-section allows for isolation of the contributions of the bonds along different directions. By exploiting the linear polarization of the synchrotron radiation and by orienting the sample with the surface normal either parallel or perpendicular to the electric vector of the impinging x-rays, it is possible to probe either the out-of-plane or in-plane atomic bonds. This analysis provides direct information about strain in well-oriented thin films such as MnSi on Si(111) and could be used to investigate other systems where the local strain at the interface affects their electronic and magnetic properties as in the case of superconducting lattices\cite{FogelPRL2001,TangNatPhys2014} or heterostructures incorporating topological insulators.\cite{VobornikNanoLett2014,LiPRL2015} In the following we will describe our results on differently prepared MnSi thin films as a function of film thickness.

Our paper is organized as follows. 
Section \ref{MnSi} reviews the state of the art on the MnSi/Si(111) system as reported in the literature. 
As the evolution of the properties of epitaxial films of MnSi depend sensitively on the film thickness in the thin film limit, earlier work provides an important point of reference for our study of thick films reported here. 
In Sec.\ \ref{Methods} we summarize the sample growth, pre-characterization of the thin films, and experimental details of our EXAFS measurements. Results of the polarization-dependent EXAFS measurements on two series of samples are presented and discussed in Sec.\ \ref{subsec:EXAFSanalysis}. Section \ref{subsec:refinement} presents the refinement results allowing for an extraction of the relative movements of the atoms in the unit cell. The somewhat surprising main result of our EXAFS measurements concerns that the Mn positions in thick MnSi films are unchanged, whereas the Si positions vary along the out-of-plane [111]-direction, whereby the orientation reverses in adjacent unit cells. Thus, for thick MnSi films, the unit cell volume is essentially unchanged from that of bulk MnSi except in the vicinity of the interface with the Si substrate. In conclusion, the mere existence of an interface appears to affect the magnetic properties of thin film MnSi strongly as discussed in Sec.\ \ref{Conclusions}.

\section{\label{MnSi}The M\lowercase{n}S\lowercase{i}/S\lowercase{i}(111) system}


In a series of seminal papers the details of epitaxial MnSi films on Si (111) have been explored in great detail for the thin film limit. 
\cite{Karhu10,Karhu11,Engelke2012,Karhu12,Wilson12,Wilson13,Li13,Menzel13,Meynell2014,Wilson14,Yokouchi15,Engelke2014}
The lattice parameter at room temperature of Si is $a_\textrm{Si} = 5.431$~\AA, whereas that of B20 MnSi is $a_\textrm{MnSi} = 4.556$~\AA. When grown on Si(111) with the epitaxial relationship Si(111) $\parallel$ MnSi(111) and $[11\bar{2}]$Si $\parallel$ $[1\bar{1}0]$MnSi [cf., Fig.\ \ref{fig:structure}(a)], the lattice mismatch between MnSi(111) and Si(111), as determined by $( a_\textrm{MnSi}  \cos30^\circ - a_\textrm{Si}) / a_\textrm{Si}$, is $-$3.0\%.\cite{Karhu10,Karhu12} This tensile strain represents a comparably large value and the formation of a dislocation network can be expected, still resulting in some residual strain. The defects have been identified as misfit edge dislocations with a Burgers vector of $b = a_\textrm{MnSi} \cdot [\bar{a}10]/2$.\cite{Karhu10,Karhu12}

In-plane and out-of-plane strain has been studied by various groups using, e.g., transmission electron microscopy (TEM) plane-view selected-area diffraction patterns\cite{Karhu10,Karhu12} and density functional theory calculations.\cite{Geisler2013} The studies focused on reactively grown MnSi layers and did not cover stoichiometrically grown films. The reactively grown films show an interesting increase of the out-of-plane strain with film thickness up to 70~\AA, above which the strain reduces again. The in-plane strain, on the other hand, smoothly reduces as the films are grown thicker.\cite{Karhu12}
In MBE, MnAs films can be grown up to a thickness of $\sim$500~\AA.\cite{Zhang_MnSi_2016}

\begin{figure}
	\includegraphics[width=0.8\columnwidth]{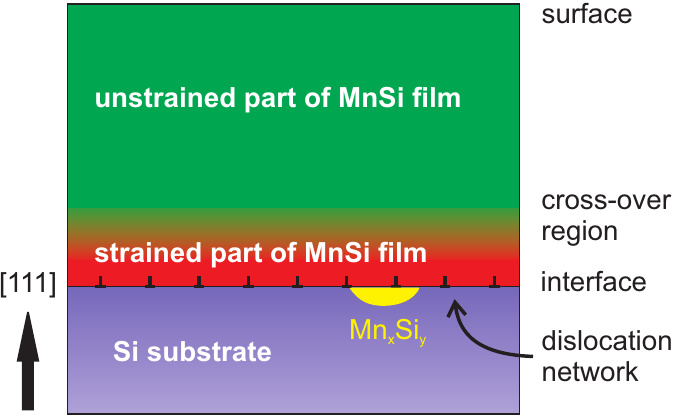}
	\caption{Illustration of the structural properties of the MnSi/Si(111) system, summarizing the relevant literature reports and our own findings. Reactive growth is accompanied by the formation of secondary phases in the substrate, right at the interface. The lattice mismatch between Si and MnSi is accommodated via the formation of a dislocation network. The residual strain affects the film close to the interface, up to a thickness of $\sim$100~\AA, resulting in a thickness-dependent magnetic transition temperature. The unstrained part of the film, reaching up to $\sim$500~\AA, is characterized by a constant transition temperature.}
	\label{fig:layers}
\end{figure}

The influence of strain on the magnetic transition temperature has been measured as a function of film thickness (see Fig.~6 in Ref.~\onlinecite{Karhu10}). At 70~\AA, where the out-of-plane strain is at its maximum, the measured magnetic transition temperature $T_c$ coincides with the bulk value for MnSi. Above 100~\AA, the value saturates at about 15~K above the bulk value. 
Figure \ref{fig:layers} summarizes the structural properties of MnSi thin films on Si(111).
In our films, which are grown by a combination of reactive seed layer formation and stoichiometric deposition, much higher transition temperatures were found.\cite{Zhang_MnSi_2016}

Given further that the electronic properties of Si do not appear to be essential for the formation of the skyrmions, the most prominent effect of the interface appears to be the occurrence of strain, affecting the delicate balance of the energy terms stabilizing the skyrmion phase. However, in order to clarify the precise mechanism causing the enhanced magnetic transition temperature detailed measurements of the structural properties of epitaxial MnSi film in the thick film limit are required as reported below.


\section{\label{Methods}Materials System and Experimental Methods}


\subsection{\label{MBE}Sample preparation and characterization}

\begin{figure}
 \includegraphics[width=1\columnwidth]{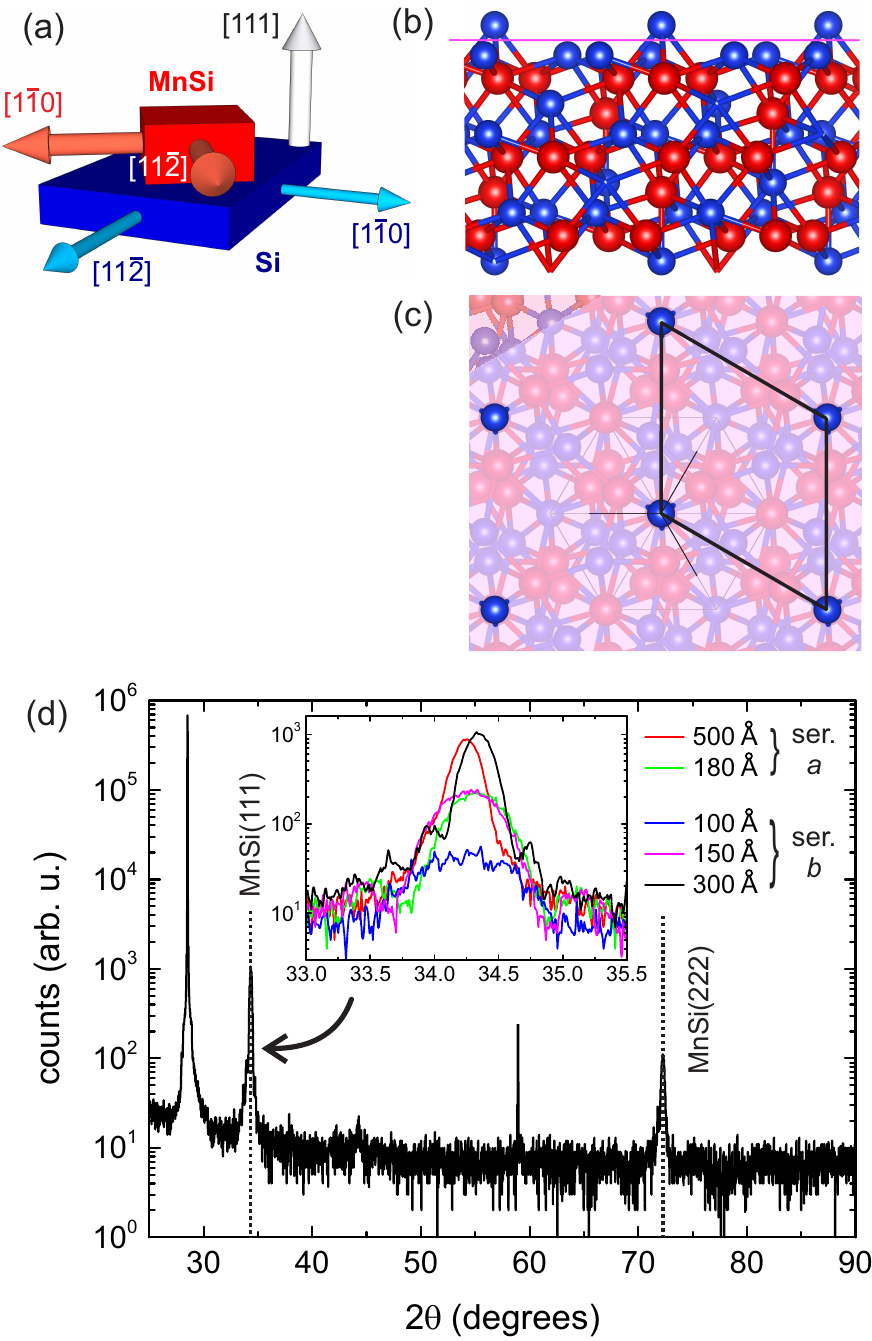}
\caption{Crystal structure and structural properties. (a) Illustration of the epitaxial relationship of B20 MnSi on Si(111): MnSi[1$\bar{1}$0]$\parallel$Si[11$\bar{2}$] and MnSi(111)$\parallel$Si(111).
(b) Side-view looking down the [$\bar{1}$10] direction of MnSi, where Mn atoms are shown in red and  Si atoms in blue. The (111) plane is indicated by a horizontal pink line.
(c) In the top view of the (111) surface only the Si atoms above the purple (111) lattice plane are shown bright. This $(\sqrt{30} \times \sqrt{30})$R30$^\circ$ mesh is closely matched to Si(111).
(d) X-ray diffraction spectrum of both series of MnSi films on Si(111). An overview $2 \theta$-$\omega$ scan for the 300-\AA-thick unbuffered sample (series $b$) shows the MnSi film and Si substrate peaks. The inset shows the MnSi(111) film peaks for the entire series.}
\label{fig:structure}
\end{figure}

Epitaxial MnSi(111) thin films were grown by MBE on Si$(111)$ substrates
measuring 10 $\times$ 12~cm$^2$.
The Balzers MBE system has a base pressure of $5 \times
10^{-10}$~mbar and is equipped with electron-beam evaporators and
effusion cells. Flux control is achieved via cross-beam mass
spectrometry. Prior to loading, the Si wafers were first degreased,
followed by etching in hydrofluoric acid and oxidation by
H$_2$O$_2$. Annealing at 990$^{\circ}$C, and growing a Si buffer
layer, leads to the $(7 \times 7)$ reconstruction, as confirmed by
reflection high energy electron diffraction (RHEED). The sample is
then cooled down to room-temperature and $\sim$3 monolayers (MLs) of Mn
were deposited before they are reacted with the Si surface at an
elevated temperature of $\sim$250-300$^\circ$C. (1~ML
corresponds to $7.82 \times 10^{14}$ atoms/cm$^2$.) This leads to the
formation of an epitaxial MnSi seed layer. The MnSi layer has a
$(\sqrt{30} \times \sqrt{30})$R30$^\circ$ structure [Fig.\ \ref{fig:structure}(c)], as determined
by RHEED. The subsequent MnSi growth is by the stoichiometric supply
of Mn and Si. The growth proceeds up to a thickness of roughly
500\,\AA\ without any signs of the formation of a secondary phase.
It is worth noting that the surface of the seed layer prepared in
this way (reactive epitaxy) is not perfect.\cite{Geisler2012_STM} In fact, choosing the described Mn thickness and annealing conditions assures that the
surface is as flat as possible. The in-vacuo study of freshly grown
MnSi layer by scanning tunneling microscopy (STM) showed a large variation in surface morphology,\cite{Kumar2004}
however, its effect on strain in the subsequent layer is not known
in detail.
Figure \ref{fig:layers} summarizes the structural properties of the MnSi/Si(111) system as reported in the literature, and also includes our own findings in the thick film limit.

Two groups of samples were grown, namely with (series $a$) and without (series $b$) a Si buffer layer. Both series include different thicknesses, as confirmed by x-ray reflectivity, of 100, 150, 180, 300, and 500~\AA, thereby covering the range from a couple of ML thick seed layers to the maximum achievable B20 phase film thickness. All films were characterized by XRD [Fig. \ref{fig:structure}(d)]. In order to protect the films from oxidation, a thin amorphous silicon cap was added to some of them.

\begin{figure}
 \includegraphics[width=1\columnwidth]{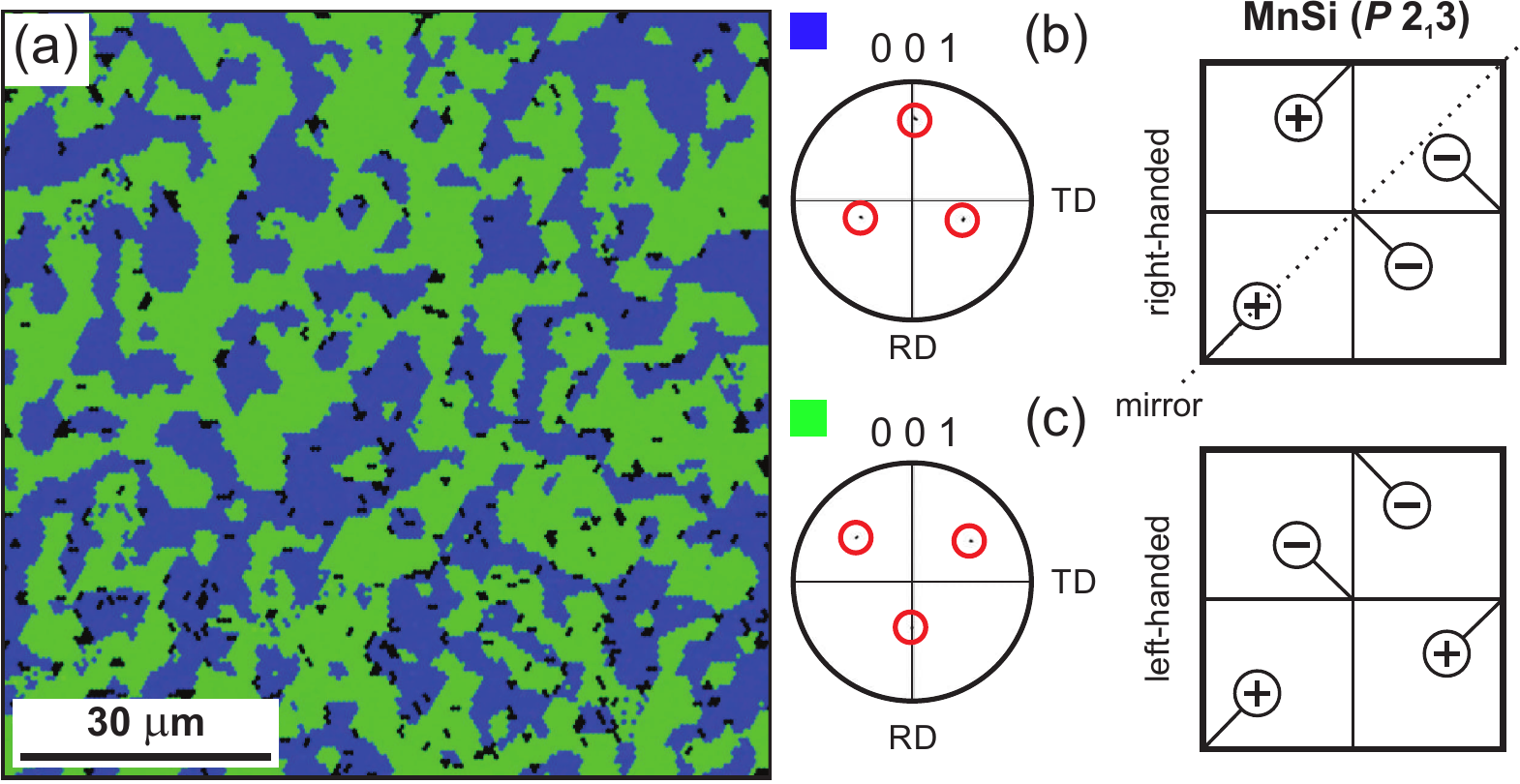}
\caption{(a) EBSD orientation map of a 500-\AA-thick MnSi film showing the in-plane orientations of the left- and right-handed form of MnSi. (b,c) Pole figures for the right-handed domains (blue) and left-handed domains (green). The red circles highlight the positions of the diffraction spots. 
As $P2_13$ MnSi has rotational symmetry of order three about the [111] direction (film normal), the two domains are essentially rotations about [111] of 60$^\circ$ relative to each other.  This is equivalent to keeping the crystal axes unrotated but having the other chirality of the crystal, as illustrated in the corresponding unit cell configurations on the right.}
\label{fig:EBSD}
\end{figure}

We found chiral domains in the films and visualized them using electron backscatter diffraction (EBSD) in an JEOL 6500F scanning electron microscope (SEM) fitted with an EBSD phosphor screen detector and CCD camera. The EBSD map of the 500-\AA-thick MnSi film [Fig.~\ref{fig:EBSD}(a)] shows that the chiral domains, which have a correlation length on the order of 10~$\mu$m, are equally distributed.

\begin{figure}
 \includegraphics[width=1\columnwidth]{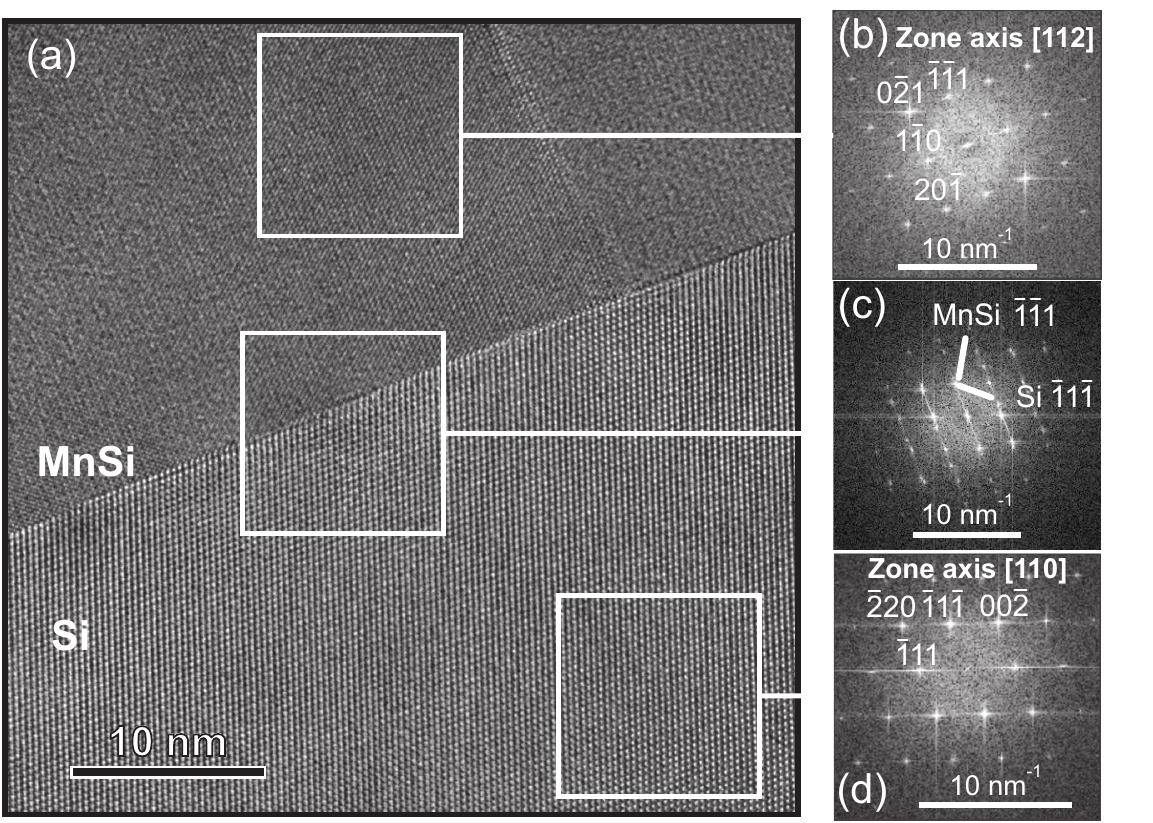}
\caption{(a) Cross-sectional high-resolution TEM image of MnSi/Si(111). Power spectra of (b) the MnSi film, (c) MnSi/Si(111) interface, and (d) Si substrate.}
\label{fig:TEM}
\end{figure}

The sample for cross-sectional TEM  
was prepared by mechanical polishing followed by Ar ion-beam
milling. The high-resolution TEM  phase-contrast image [Fig.\
\ref{fig:TEM}(a)] indicates that the MnSi film is a highly ordered,
epitaxial structure on the Si(111) substrate, and that the interface
between substrate and film is sharp. The power spectra observed at
different locations of the film [example shown in Fig.\
\ref{fig:TEM}(b)] are identical and can be indexed to the B20 cubic
MnSi phase (space group $P2_13$, $a = 0.4556$~nm) with a zone axis
of [112]. 
The Si(111) substrate power spectrum is indexed as the diamond space group $Fd\bar{3}m$ and the zone axis is identified as
[110].

Another important fact is that the power spectra obtained at different locations of the Si substrate do not show a significant or dominant presence of other Mn-Si phases in the substrate.
Minor cases of inclusions were detected by TEM, as reported previously.\cite{Zhang_MnSi_2016}
The power spectrum calculated at the
interface between substrate and film [Fig.\ \ref{fig:TEM}(d)] is a
combination of the power spectra from Si(111) and MnSi and it shows
that ($\bar{1}\bar{1}1$) planes of MnSi and ($\bar{1}1\bar{1}$)
planes of the Si substrate are parallel. From these observations,
the epitaxial relationship between MnSi film and Si substrate can be
written as MnSi\{111\}$\langle$112$\rangle$$\parallel$ Si\{111\}$\langle$110$\rangle$.\\

\subsection{\label{EXAFS-meas}EXAFS measurements and analysis}

X-ray absorption near edge structure (XANES) and EXAFS spectra were
collected at the Mn $K$ edge ($\sim$6539~eV) of
the thin film samples at room temperature on beamline B18 at the
Diamond Light Source. Bulk MnSi in powder form and a Mn foil
were measured as references. The powdered MnSi bulk was pressed into
a pellet with the optimized quantity for measurements in
transmission mode.

A nine-element solid-state Ge detector with
digital signal processing for fluorescence detection was used to measure the thin
films. The polarization dependence arises from the different geometries employed for data collection. Measurements were performed in two orientations of the films with respect to the polarization vector of the linearly polarized synchrotron radiation, so that this vector is aligned either parallel or perpendicular to the sample surface, i.e., at an 
incident angle of 10$^\circ$ [normal, with  electric field vector {\textbf{E}} of the
x-rays nearly parallel to the MnSi(111) plane] and 80$^\circ$ [grazing, with {\textbf{E}} nearly
perpendicular to the MnSi(111) plane],  c.f.\ Fig. \ref{fig:structure}(a). 
All spectra
were acquired in quick-EXAFS mode using the branch of Pt-coated collimating and focusing mirrors, a Si(111) double-crystal monochromator, and a
pair of harmonic rejection mirrors. The energy range of the scans allowed us
to extract information in the extended region up to $k$  $\approx$ 14~\AA$^{-1}$. 
An average of 20 and 30 scans were recorded per sample for each
orientation.


EXAFS spectra were processed and analyzed using  various tools of
the {\textsc{iffefit}} XAFS package.\cite{Ravel2005} This involved
preliminary reduction of the EXAFS raw data, background removal of
the x-ray absorption data $\mu(E)$, conversion of $\mu(E)$ to the EXAFS modulation function 
$\chi(k)$, with $k$ the wave number of the photoelectron, normalization and weighting scheme; all of them performed
with {\textsc{autobk}} and {\textsc{athena}}. EXAFS data analysis
and fitting on all references and samples were performed in
{\textsc{artemis}}, making use of models based on crystallographic
information obtained from the ICSD database. The atomic clusters
used to generate the scattering paths for fitting were generated
with {\textsc{atoms}}.\cite{Ravel2001}

\section{\label{Results}Results}

\subsection{\label{subsec:EXAFSanalysis} Bond-distance analysis}

Polarization-dependent EXAFS is used to obtain
anisotropic static strain in thin films, as it yields the in-plane
(ip) and out-of-plane (oop) average bond distances and nearest
neighbors of a given atom in a crystalline and well-oriented
sample. At the Mn $K$ edge of the MnSi thin films such measurements
provide information about the anisotropic local structure of the Mn
atoms.

\begin{figure}
 \includegraphics[width=1.0\columnwidth]{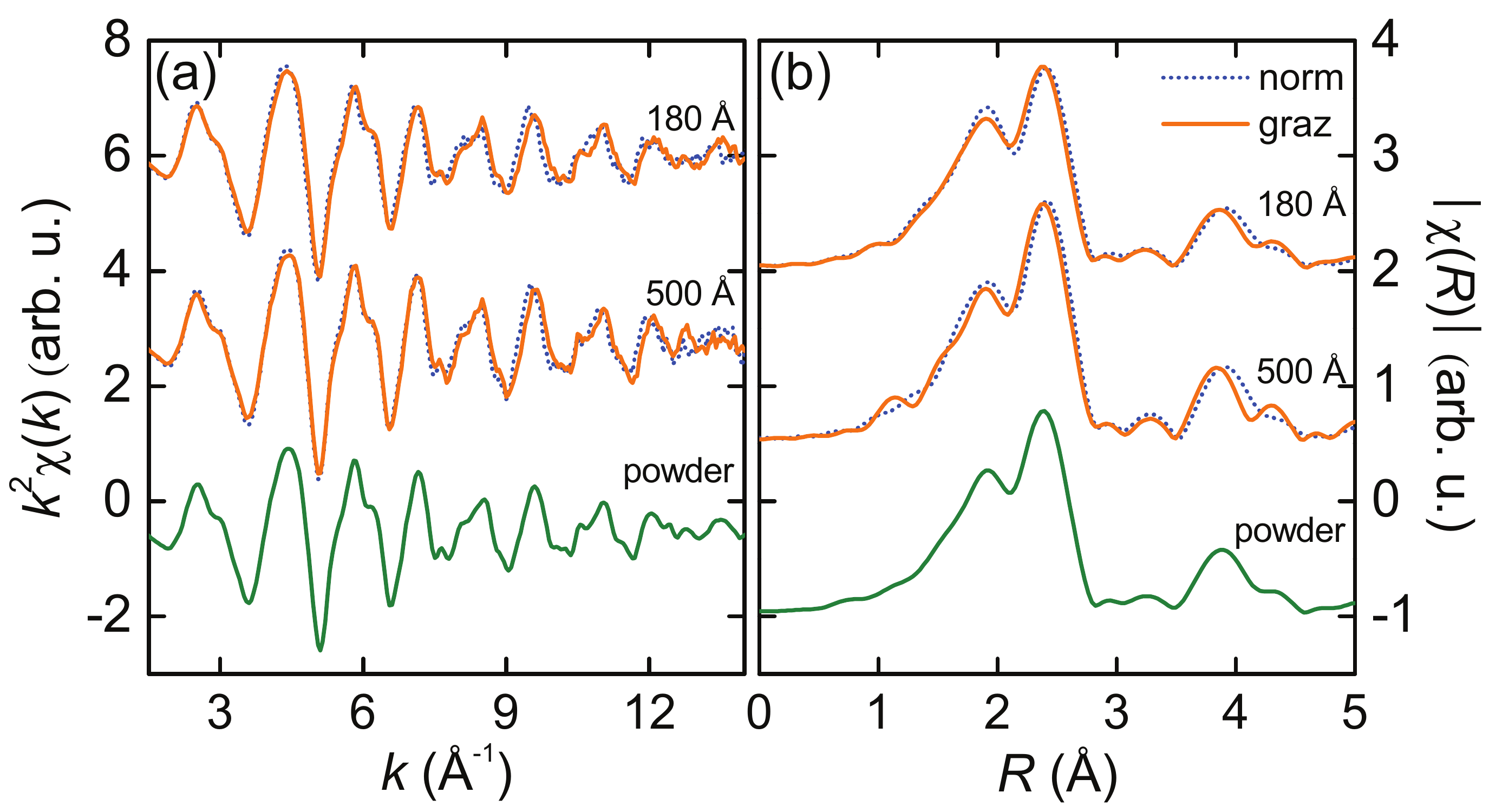}
\caption{(a) EXAFS signal at the Mn $K$ edge of the MnSi thin films
with Si buffer layer measured at in-plane (solid line) and out-of-plane (dotted
line) orientations. The spectrum for bulk MnSi is included for
comparison. (b) Fourier transform of the EXAFS signal in (a). Curves
have been vertically shifted for clarity. } \label{fig:EXAFS_Si}
\end{figure}

\begin{figure}
 \includegraphics[width=1.0\columnwidth]{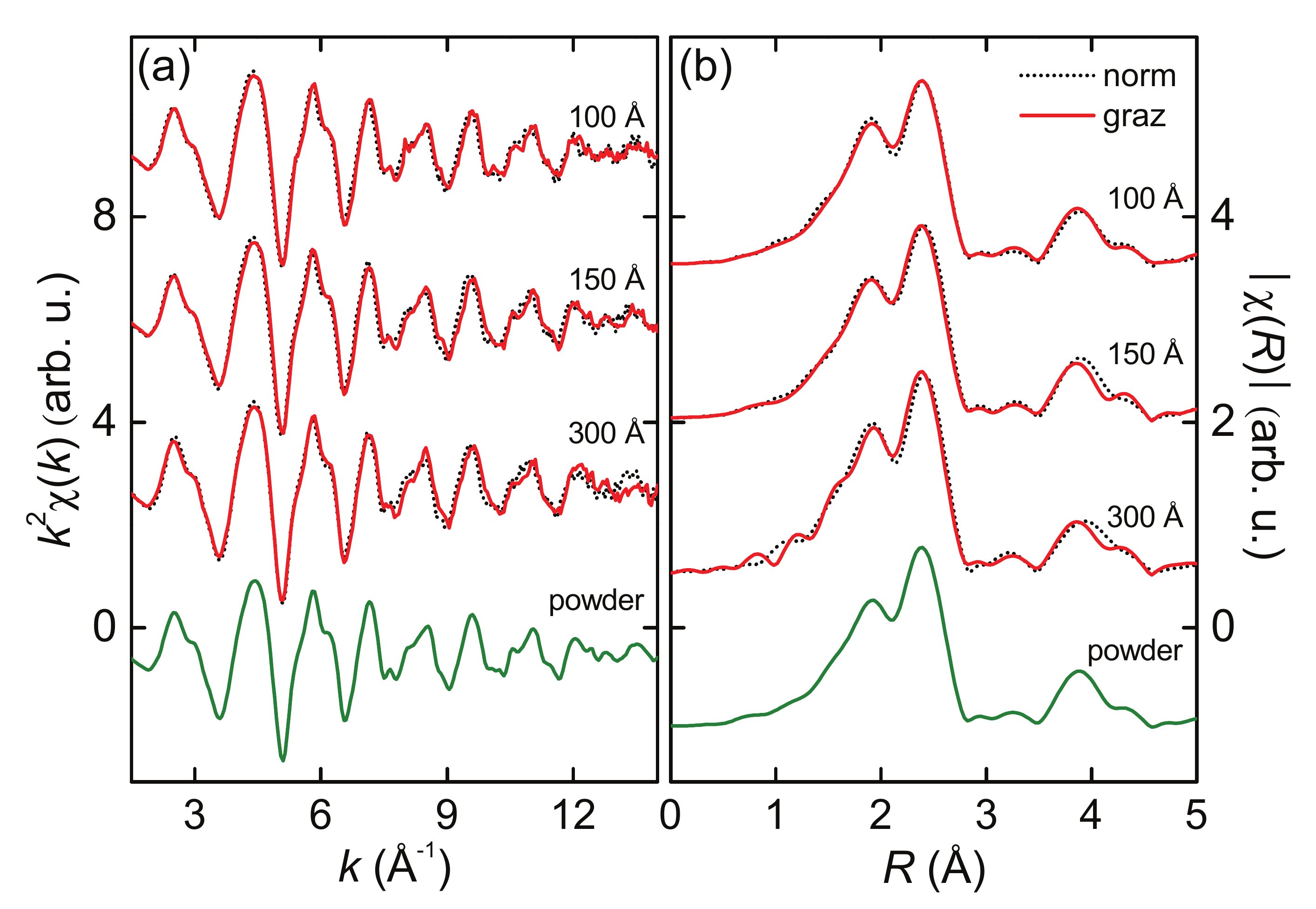}
\caption{(a) EXAFS signal at the Mn $K$ edge of the MnSi thin films
without Si buffer layer measured on the  in-plane (solid line) and out-of-plane
(dotted line) geometries. The spectrum for bulk MnSi is included for
comparison. (b) Fourier transform of the EXAFS signal in (a). Curves
have been vertically shifted for clarity.} \label{fig:EXAFS_NOSi}
\end{figure}

Figures \ref{fig:EXAFS_Si} and \ref{fig:EXAFS_NOSi} show the raw
$\chi(k)$ EXAFS signal and module of the Fourier transform (FT) of
the thin films at normal and grazing geometries performed over a
3-13~{\AA}$^{-1}$ $k$-range using a Hanning window function, and
$\Delta k=2$ {\AA}$^{-1}$. All plots are performed using a $k^2$
weight. Data obtained for the powdered bulk MnSi is also included in
these figures as reference. The shape and intensity of the spectra
for thin films is, in general, similar to the one for bulk MnSi,
with slight differences that might be related with the strain
effects that we are looking for. The comparison between the normal and grazing incidence spectra for each sample shows
subtle differences, which can be quantified by fitting the EXAFS
signals.

\begin{figure}
 \includegraphics[width=1.0\columnwidth]{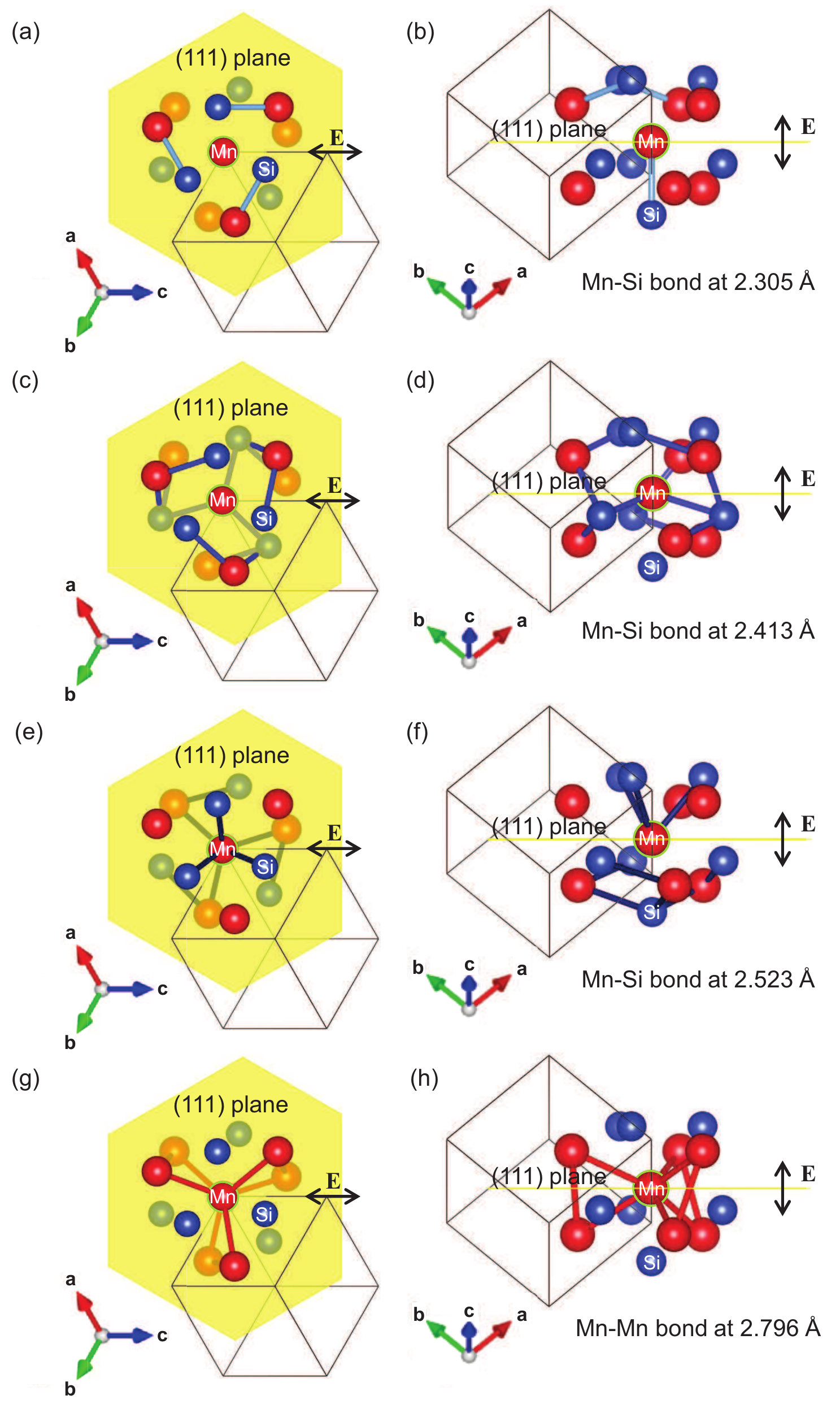}
\caption{Illustration of the bonds and geometry with respect to the
(111) lattice plane of the MnSi crystal and {\textbf{E}}-vector of the x-rays. A representative Mn absorbing atom on one of the MnSi(111) planes is highlighted and its bonds for the first and
second coordination shells are depicted: (a,b) Mn-Si bond at 2.305 \AA, (c,d) Mn-Si bond at 2.413 \AA, (e,f) Mn-Si bond at 2.523 \AA, and (g,h) Mn-Mn bond at 2.796 \AA. Left (right) panel shows
the orientation of the bonds with respect to {\textbf{E}} in
normal (grazing) geometry. The crystallographic planes and directions of the MnSi thin films are illustrated in Fig.\ \ref{fig:structure}(a). All bonds and geometries are generated using crystallographic information from ICSD 71830.\cite{Jorgensen1991}
} \label{fig:MnSi_geom}
\end{figure}

According to crystallographic information available for B20 MnSi
in ICSD (71830),\cite{Jorgensen1991} the first coordination
shell of Mn atoms in this structure is composed of seven Si
neighbors (one at 2.305~\AA, three at 2.413~\AA, and three at
2.523~\AA) and a second shell with six Mn neighbors at 2.796~\AA.
However, for the oriented thin films, given that the x-ray absorption is polarization-dependent, it is necessary to define an
effective coordination number\cite{Bunker10} as
\begin{equation}
\centering N^* = 3 \sum_{i=1}^N \cos^2 \alpha_i  \, ,
\label{eq:Neff}
\end{equation}
\noindent where $\alpha_i$ is the angle between the electric-field vector  {\textbf{E}}  and the direction of the bond
between the Mn atom and the neighbor $i$ of a given coordination
shell with $N$ atoms. Note that Eq.~(\ref{eq:Neff}) is only valid for polarization-dependent measurements at the $K$ edge.\cite{Bunker10} Taking the bonds from the
first and second coordination shells for MnSi for each geometry
(grazing and normal), as illustrated in Fig.~\ref{fig:MnSi_geom},
$N^*$ for each scattering path is calculated and listed in Table
\ref{tab:EXAFSFits}.

\begin{figure}
 \includegraphics[width=1.0\columnwidth]{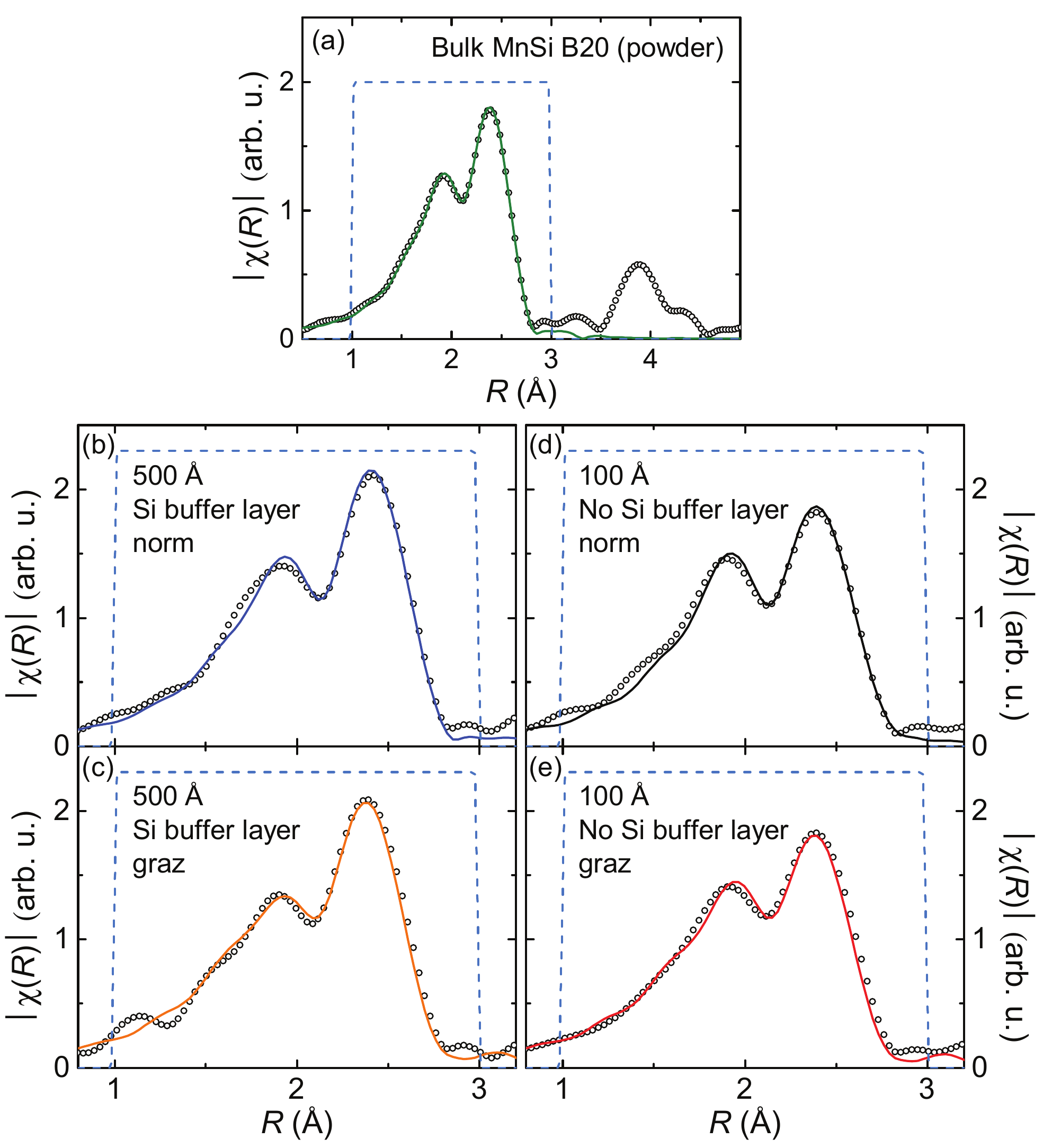}
\caption{Fourier transforms of the EXAFS signals and best fits for (a)
bulk (powder) MnSi, (b) normal geometry of a 500-\AA-thick film from
series $a$, (c) grazing geometry of the same film in (a), (d) normal
geometry of a 100-\AA-thick film from series $b$, and (e)
grazing geometry of the same film in (d).}
\label{fig:EXAFS_Fits}
\end{figure}

Fits of the Mn $K$-edge EXAFS signal were performed in $R$-space in
a range of 1-3~\AA\ using a Hanning window function, so that it
covered the first two coordination shells of the Mn atoms (Fig.~\ref{fig:EXAFS_Fits}). The parameters fitted for the bulk MnSi (powder) 
sample were a general amplitude reduction factor ($S_0^2$) and shift
in the threshold energy ($\Delta E_0$), as well as the interatomic
distance ($R$) and Debye-Waller factor ($\sigma^2$) for each
scattering path. The best fit of the EXAFS signal for bulk MnSi was
obtained using a model which groups the first two Mn-Si bonds as
four Mn-Si paths at $\sim$2.386~\AA\ (Table~\ref{tab:EXAFSFits}). This approximation is needed due to limitations of the EXAFS technique in this particular case, as the results did not allow us to distinguish between the contribution of a single atom with respect to the one of three atoms at a very close distance. For the MnSi thin films, $S_0^2$ and $\Delta
E_0$ were fixed to those values obtained for the bulk MnSi sample
($S_0^2=0.7$ and $\Delta E_0=1.03$~eV) for both geometries. Results
from these fits are listed in Table \ref{tab:EXAFSFits} and shown in
Fig.~\ref{fig:EXAFS_Fits}. 
The Mn-Si-$\textsl{1}$, $\textsl{2}$, and $\textsl{3}$ bonds refer to the first, second, and third coordination shells, respectively.

\begin{table*}[t]
\caption{Structural parameters obtained from the fit of the normal
(norm) and grazing (graz) data of the thin film samples with
different thickness $t$ of MnSi. The effective models for a non-oriented
MnSi system (aver), an oriented MnSi system with {\textbf{E}}
$\parallel$ [111] (graz) and an oriented MnSi system with
{\textbf{E}} $\perp$ [111] (norm) are listed as reference.
Results obtained for the bulk (powder) MnSi reference are shown for
comparison. Effective coordination number, $N^*$, interatomic
distance, $R$, and Debye-Waller factor, $\sigma^2$, for each path.
The uncertainty in $R$ is $\pm$0.006~\AA\ and that in $\sigma^2$ is
$\pm$20\%.}\begin{ruledtabular}
\begin{tabular}{lcccccccccccccc}
 & $t$ & Geom & $N^*_\textrm{Mn-Si-\textsl{1}}$ & $R_\textrm{Mn-Si-\textsl{1}}$ & $\sigma^2_\textrm{Mn-Si-\textsl{1}}$ & $N^*_\textrm{Mn-Si-\textsl{2}}$ & $R_\textrm{Mn-Si-\textsl{2}}$ & $\sigma^2_\textrm{Mn-Si-\textsl{2}}$ & $N^*_\textrm{Mn-Si-\textsl{3}}$ & $R_\textrm{Mn-Si-\textsl{3}}$ & $\sigma^2_\textrm{Mn-Si-\textsl{3}}$ & $N^*_\textrm{Mn-Mn}$ & $R_\textrm{Mn-Mn}$ & $\sigma^2_\textrm{Mn-Mn}$\\
 & (\AA) & & & (\AA) & (\AA$^2$) & & (\AA) & (\AA$^2$) & & (\AA) & (\AA$^2$)& & (\AA) & (\AA$^2$)\\
\hline
Model\footnotemark[1] & - & aver & 1 & 2.305 & - & 3 & 2.413 & - & 3 & 2.523 & - & 6 & 2.796 & -\\
Model\footnotemark[2] & - & norm & - & - & - & 4.12 & 2.413 & - & 1.86 & 2.523 & - & 6.98 & 2.796 & -\\
Model\footnotemark[2] & - & graz & 3 & 2.305 & - & 0.76 & 2.413 & - & 5.27 & 2.523 & - & 4.04 & 2.796 & -\\
Bulk & - & - & - & - & - & 4 & 2.383 & 0.008 & 3 & 2.525 & 0.023 & 6 & 2.772 & 0.006\\
\hline
\multirow{4}{*}{Series $a$} & 500 & norm & - & - & - & 4.12 & 2.377 & 0.006 & 1.86 & 2.503 & 0.007 & 6.98 & 2.791 & 0.006\\
& 500 & graz & 3.76 & 2.405 & 0.016 & - & - & - & 5.27 & 2.441 & 0.014 & 4.04 & 2.774 & 0.003\\
& 180 & norm & - & - & - & 4.12 & 2.380 & 0.005 & 1.86 & 2.536 & 0.006 & 6.98 & 2.801 & 0.008\\
& 180 & graz & 3.76 & 2.388 & 0.010 & - & - & - & 5.27 & 2.467 & 0.020 & 4.04 & 2.773 & 0.004\\
\hline
\multirow{6}{*}{Series $b$} & 300 & norm & - & - & - & 4.12 & 2.372 & 0.005 & 1.86 & 2.502 & 0.007 & 6.98 & 2.789 & 0.007\\ 
& 300& graz & 3.76 & 2.440 & 0.027 &  - & - & - & 5.27 & 2.418 & 0.010 & 4.04 & 2.773 & 0.003\\
& 150 & norm & - & - & - & 4.12 & 2.383 & 0.005 & 1.86 & 2.536 & 0.005 & 6.98 & 2.792 & 0.007\\
& 150 & graz & 3.76 & 2.411 & 0.010 &  - & - & - & 5.27 & 2.445 & 0.021 & 4.04 & 2.772 & 0.003\\
& 100 & norm & - & - & - & 4.12 & 2.382 & 0.004 & 1.86 & 2.549 & 0.005 & 6.98 & 2.789 & 0.008\\
& 100 & graz & 3.76 & 2.413 & 0.008 & - & - & - & 5.27 & 2.443 & 0.023 & 4.04 & 2.773 & 0.003\\
\end{tabular}
\end{ruledtabular}
\label{tab:EXAFSFits} \footnotetext[1]{Bond distances extracted from
ICSD 71830.\cite{Jorgensen1991}} \footnotetext[2]{Effective
coordination numbers calculated using Eq.~(\ref{eq:Neff}) and bonds
obtained from ICSD 71830.\cite{Jorgensen1991}}
\end{table*}

\begin{figure}
 \includegraphics[width=1.0\columnwidth]{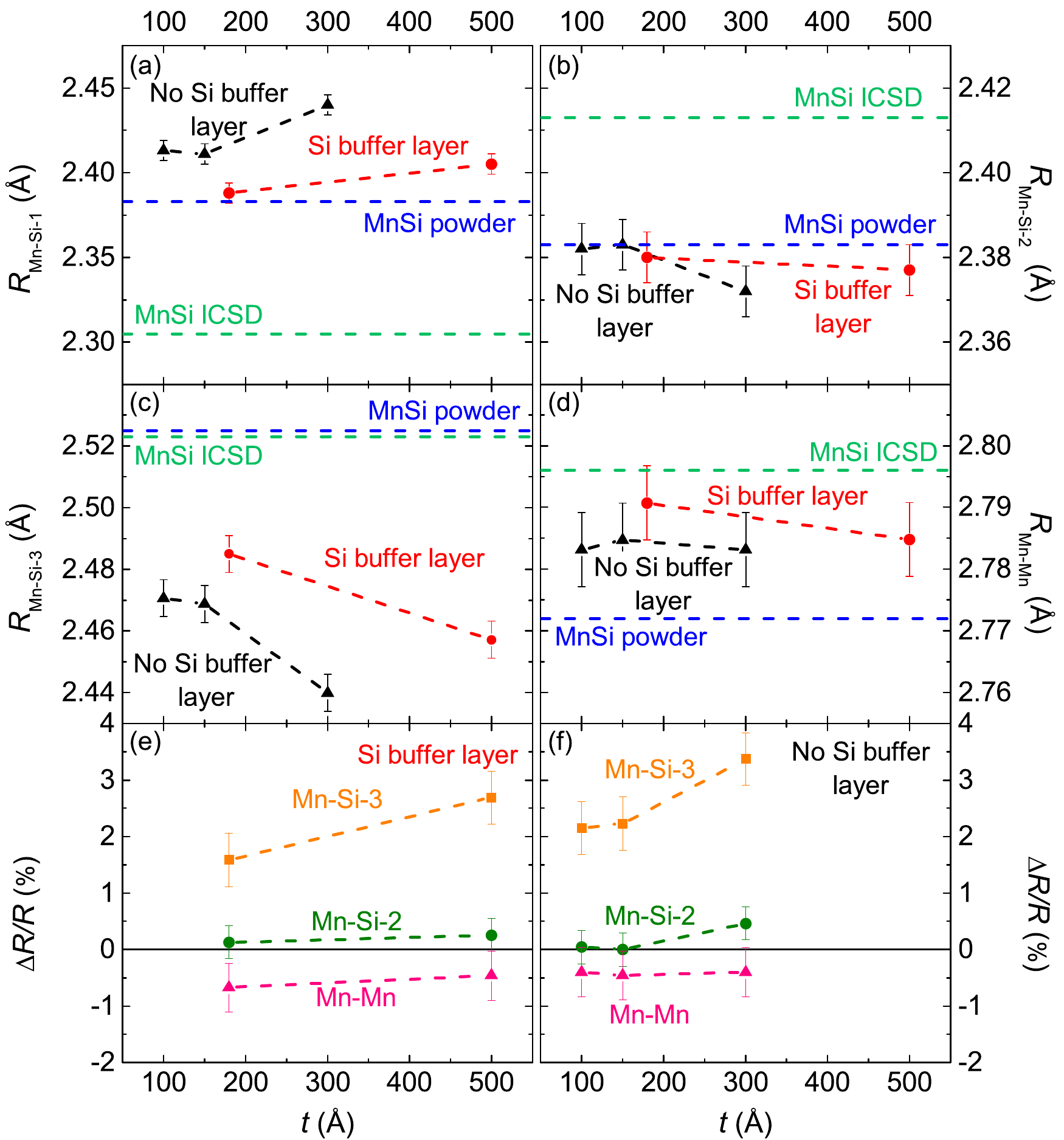}
\caption{(a-d) Bond lengths extracted from the polarization-dependent EXAFS fits for the first and second coordination shell of Mn atoms in the MnSi thin films, plotted as a function of film thickness. The
horizontal, dashed lines on each graph give the reference values of the
result obtained for bulk (powder) MnSi for that particular bond and the bond
distance from the crystallographic model (ICSD 71830). (e,f) Depiction of the results of $\Delta R/R$ for each bond on the thin films with and without a Si buffer layer, respectively.} \label{fig:MnSi_bonds}
\end{figure}

From Table~\ref{tab:EXAFSFits} and Fig.~\ref{fig:MnSi_geom} we
can conclude that there are bonds that contribute more to the EXAFS
signal in one particular geometry. In fact, the Mn-Si-$\textsl{1}$ bond
only contributes to the signal in grazing geometry, whereas
Mn-Si-$\textsl{2}$  contributes mostly to that in normal incidence. The amplitude of the latter in grazing geometry is so low that
it has no effect on the fit. Results of the bond length $R$ obtained
for these two bonds are plotted for all thin films in Figs.~\ref{fig:MnSi_bonds}(a) and \ref{fig:MnSi_bonds}(b), respectively.
The values obtained for bulk (powder) MnSi for each bond and those from the
crystallographic model (ICSD 71830\cite{Jorgensen1991}) are marked as reference on each plot.

$R_\textrm{Mn-Si-\textsl{1}}$ and its Debye-Waller factor increase with thickness for both sample series. This bond is longer for samples without Si buffer (series $b$) than those with buffer. For the 300-\AA-thick sample, this bond length is at its maximum and the Debye-Waller factor is rather large, which suggests a large degree of disorder and/or distortion of this out-of-plane bond. Values for $R_\textrm{Mn-Si-\textsl{2}}$ vary less within and between the sample series and are found to be very close to the value obtained for bulk MnSi. This result may indicate less in-plane distortion and/or strain in the thin films.

For Mn-Si-$\textsl{3}$ and Mn-Mn bonds, given that they contribute to
both grazing and normal geometry, we have calculated a weighted
average from the values obtained from the EXAFS fits. Resulting bond lengths are plotted in Figs.~\ref{fig:MnSi_bonds}(c) and \ref{fig:MnSi_bonds}(d), respectively. $R_\textrm{Mn-Si-\textsl{3}}$ for both sample series increases for decreasing thickness.
There seems to be a slight increase in $R_\textrm{Mn-Mn}$ for both sample series, but it is within the error bar.

In order to analyze these results, we define the parameter
\begin{equation}
\frac{\Delta R}{R} \equiv \frac{R_{\textrm{powder}}-R_{\textrm{film}}}{R_{\textrm{powder}}}\times 100 \%  \, ,
\label{eq:strain}
\end{equation}
\noindent
where $R_{\textrm{film}}$ is the bond length obtained from  EXAFS fits for a given bond on a thin film sample and $R_{\textrm{powder}}$ that obtained for the same bond for bulk (powder) MnSi. $\Delta R/R$, expressed in \%, provides quantitative information about the change in bond length of the films compared to that of the bulk (powder) sample, and it is proportional to the strain in the system. Results of $\Delta R/R$ obtained for each bond are plotted in Figs.~\ref{fig:MnSi_bonds}(e) and \ref{fig:MnSi_bonds}(f) for the samples with and without Si buffer layer, respectively.

For films with Si buffer layer, $\Delta R/R$ for each bond does not vary significantly with film thickness [cf.\ Fig.~\ref{fig:MnSi_bonds}(e)], whereas for samples of series $b$ without a buffer, the value increases with increasing film thickness [cf.\ Fig.~\ref{fig:MnSi_bonds}(f)]. $\Delta R/R$ for Mn-Si-$\textsl{2}$ bonds on both sample series is negligible, which suggests low in-plane strain. Larger values of $\Delta R/R$ are found for Mn-Si-$\textsl{3}$ bonds, which, according to the model used for the analysis of the polarization dependence, is related to out-of-plane strain. It reaches 3.5\% in samples without Si buffer, and 2.7\% in those with a Si buffer.

These out-of-plane strain values, when taken as unit cell strain in an epitaxial film, are very large, whereas the in-plane strain is negligible. Interestingly, in this picture, the out-of-plane lattice constants of the films appear smaller than that of the bulk, meaning that the films should be compressively strained. This behavior is in contrast to previous reports in which tensile strain was found both in-plane\cite{Wilson14} and out-of-plane, using XRD and TEM, leading to an increase of the unit cell volume.\cite{Karhu10}
In Ref.\ \onlinecite{Karhu10}, it was further shown that the critical temperature $T_\mathrm{c}$ of the films is a linear function of the film thickness up to $\sim$10~nm, above which the $T_\mathrm{c}$ saturates at $\sim$42~K (for films up to 22~nm thick).
For compressively strained MnSi, e.g., in hydrostatic pressure studies of bulk crystals, a decrease of $T_\mathrm{c}$ with increasing pressure was reported.\cite{Pfleiderer04}
In fact, above 14.6~kbar long-range magnetic order is fully suppressed.\cite{Pfleiderer04}
Recent hydrostatic pressure studies on MnSi thin films on Si(111) have shown that this critical pressure is enhanced to 31~kbar.\cite{Engelke2014}

Using the elastic constants of MnSi of $c_{11} = 3.21 \times 10^{12}$~dyn/cm$^2$, $c_{12} = 0.85\times 10^{12}$~dyn/cm$^2$, and $c_{44} = 1.26\times 10^{12}$~dyn/cm$^2$ (all at 6.5~K),\cite{Petrova2011} we obtain a bulk modulus for this cubic system of $K = (c_{11} + 2 c_{12})/3 = 1.64\times10^{12}$~dyn/cm$^2$ = 164~GPa and a shear modulus of $\mu = c_{44}  
$ = 126~GPa. Young's modulus can then be calculated from $E = 9 K \mu / (3 K + \mu) = 301$~GPa.\cite{Thorne}
With the tensile stress $\sigma = E \epsilon$, where $\epsilon = \Delta L / L$ is the strain, we obtain $\sigma / E = $1.46~GPa/(301~GPa) $\approx 0.005$ (3.1~GPa/(301~GPa) $\approx 0.01$), i.e., above a strain value of 0.5\% (1.0\%) magnetic long-range order, and thus the skyrmion state, are suppressed in bulk crystals (films).

Interestingly, for the films studied here, a $T_\mathrm{c}$ of $\sim$46.6~K (series $a$) and $\sim$44~K (series $b$), respectively, was observed, roughly consistent with the previously reported values for thick, epitaxial MnSi films ($>$10~nm).\cite{Karhu10}
As these values for the transition temperature are in fact equal or even slightly higher than the values for the tensile strained films, strain does not seem to have a straightforward and direct effect on magnetic ordering in MnSi films (in the thick film limit).
In order to shed more light on the apparent discrepancy, we performed a structure refinement of the crystal structure based on the EXAFS results.

\subsection{\label{subsec:refinement}Lattice structure analysis}

One of the main conclusions that can be drawn from the polarization-dependent EXAFS analysis is that
the bond distances in thick, epitaxial MnSi films predominantly undergo out-of-plane changes. This finding enables us to perform a structure refinement starting from the measured bond distances.

For the structural refinement, first, the EXAFS data obtained from a (bulk) MnSi powder sample was analyzed, and a good agreement with the published ICSD data\cite{Jorgensen1991} was found.
Then, the crystalline model for bulk MnSi was used as the reference.
MnSi crystallizes in the simple cubic B20 structure with four Mn and four Si atoms in the unit cell (space group $P2_13$). The $4a$ Wyckoff positions are ($u_x$, $u_x$, $u_x$), ($−u_x$+$\frac{1}{2}$, $−u_x$, $u_x$+$\frac{1}{2}$), ($−u_x$, $u_x$+$\frac{1}{2}$, $−u_x$+$\frac{1}{2}$, ($u_x$+$\frac{1}{2}$, $−u_x$+$\frac{1}{2}$, $−u_x$), with $u_\mathrm{Mn} = 0.138$ and $1 - u_\mathrm{Si} = 0.154$.\cite{Marel1998}
The Mn atom has a seven-fold coordination of Si atoms.\cite{Marel1998}
It is worth stressing again that EXAFS is only sensitive to the bond distances between atoms (not the atomic positions themselves). In our EXAFS measurements, the length of seven Si-Mn bonds and six Mn-Mn bonds were extracted. We can unambiguously associate these 14 atomic positions with the $P2_13$ space group symmetries. In other words, by setting the position of an arbitrary Mn atom as a reference, the positions of the measured 14 atoms can be determined for the MnSi bulk model. Next, it is assumed that these atoms only move along the [111] direction, which is justified from the polarization-dependent EXAFS analysis in Sec.~\ref{subsec:EXAFSanalysis}. 

\begin{figure}
 \includegraphics[width=1.0\columnwidth]{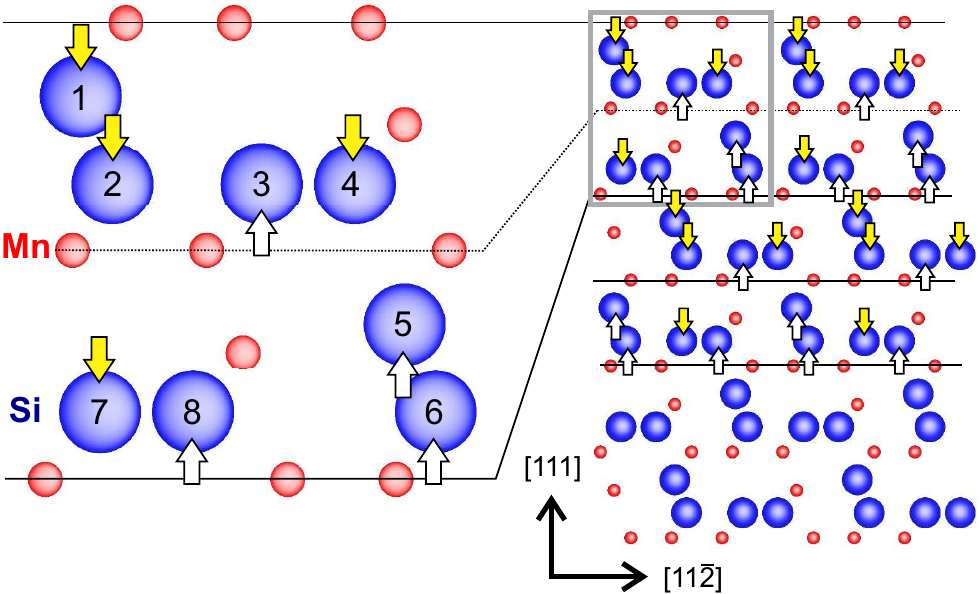}
\caption{Atomic model of MnSi in the B20 phase showing the shifts of the Si atoms (blue) with respect to the Mn atoms (red) which are unchanged.
Note that the shifts are predominantly in the [111] direction, and either up (white arrow) or down (yellow arrow).} \label{fig:Refinement}
\end{figure}

As each of these atoms shows a varied bond distance in thin films, their specific atomic positions can be determined by carrying out a data fitting procedure, giving the deviations from the Wyckoff positions summarized in Table \ref{tabrefinement}. Note that the seven positions around the Mn are now all different. The deviations from the bulk Wyckoff positions are termed $\Delta r_x$, with $x=$ Si$_i$ ($i = 1, \ldots,  8$). We formally included an eighth Si atom for illustration purposes (cf.\ Fig.~\ref{fig:Refinement}), which corresponds to a doubling of the unit cell in which one Si double-layer moves predominantly down, and the adjacent double-layer layer up, indicated by yellow and white arrows, respectively.
Interestingly, the Mn positions have not changed beyond the uncertainty in determining the shift of their position along [111] of $\pm 0.006$~\AA.

The absence of a shift of the Mn atoms means that the unit cell volume remains constant for epitaxial MnSi films in the thick film limit and we can therefore exclude (volume) strain in these films as the source for both the enhanced $T_\mathrm{c}$ and the absence of the skyrmion phase.
Further, as the Mn atoms do not move, the direct interactions between the Mn atoms are unaltered, suggesting that the balance between the exchange and the Dzyaloshinskii-Moriya interactions remains unchanged. This view is further supported by magnetometry data that show a similar behavior of MnSi thin films and bulk material.\cite{Zhang_MnSi_2016}
In a simple Heisenberg picture one would expect an increase in $T_\mathrm{c} $ only for a decreasing Mn-Mn bond distance, which we can clearly rule out.\cite{Geisler2013}
However, as the Si atoms move in films with an increased $T_\mathrm{c}$, another coupling mechanism, involving Si, may be present as well. 
For example, the change of the positions of the Si atoms could change the crystal field affecting the magnetic ions.

\begin{table*}[t]
\caption{Shifts of the positions of the Si atoms in MnSi films obtained from structural refinement assuming variations only along [111]. In comparison, the shifts of the positions of the Mn atoms are negligible. 
The labeling of the Si atoms is indicated in Fig.~\ref{fig:Refinement}.
Note that the shift of the Si7 atom can not be extracted from the EXAFS data.
The uncertainty in the determination of the shifts in the atomic positions of the Si atoms is less than $6 \times 10^{-5}$~\AA.}
\begin{ruledtabular}
\begin{tabular}{lcccccccc}
 Series & $t$ & $\Delta r_\mathrm{Si1}$ & $\Delta r_\mathrm{Si2}$ & $\Delta r_\mathrm{Si3}$ & $\Delta r_\mathrm{Si4}$ & $\Delta r_\mathrm{Si5}$ & $\Delta r_\mathrm{Si6}$ & $\Delta r_\mathrm{Si8}$ \\
 & (\AA) & (\AA) & (\AA) & (\AA) & (\AA) & (\AA) & (\AA) & (\AA) \\
\hline
\multirow{2}{*}{$a$} & 500 & $-0.0061$ & $-0.0144$ & $0.1124$ & $-0.1477$ & $0.1306$ & $0.0521$ & $0.0059$ \\
& 180 & $-0.0256$ & $-0.0637$ & 0.1279 & $-0.0784$ & 0.0708 & 0.0299 & 0.0247 \\
\hline
\multirow{3}{*}{$b$} & 300 & $-0.0319$ & $-0.0807$ & 0.1969 & $-0.2000$ & 0.1732 & 0.0660 & 0.0307 \\ 
& 150 & $-0.0233$ & $-0.0574$ & 0.1592 & $-0.1173$ & 0.1048 & 0.0429 & 0.0224\\
& 100 & $-0.0241$ & $-0.0595$ & 0.1619 & $-0.1126$ & 0.1007 & 0.0414 & 0.0232 \\
\end{tabular}
\end{ruledtabular}
\label{tabrefinement}
\end{table*}


\section{\label{Conclusions}Conclusions}
In conclusion, we have performed polarization-dependent EXAFS measurements on MnSi thin films grown on Si(111) as a function of film thickness in order to explore strain in the samples.
The investigated films are in the thick film limit, i.e., between 100 and 500~\AA, where the magnetic properties are no longer a strong function of film thickness.
Despite the high quality of the phase-pure B20 MnSi films, and the presence of a sharp interface between MnSi and Si, no unambiguous detection of skyrmions in these films has been reported. As confirmed by EXAFS analysis, the in-plane strain is negligible. The out-of-plane strain, as defined by the relative contraction of the bonds along [111], is negative and as high as 3.5\% for samples without a Si buffer layer and 2.7\% for buffered systems.
Nevertheless, the refinement of the atomic positions reveals that the change in bond lengths has different effects on Mn and Si. Whereas the positions of the Mn atoms in the crystal lattice remain unchanged, the Si atoms move along the out-of-plane [111]-direction, predominantly in opposite directions in neighboring unit cells.
The unit cell of thick MnSi films is therefore effectively unstrained.
Nevertheless, the magnetic transition temperature in the thick film limit is enhanced by $\sim$15~K as compared to the bulk, and no longer a function of film thickness, suggesting a far-reaching effect of the interface with the substrate. Another factor could be surface effects, which are effective down to a finite depth and which therefore saturate as the film is thicker than a critical thickness.
Finally, the shifted Si positions could have an indirect effect on the magnetic ordering of the Mn atoms, e.g., via the crystal field.
Therefore, more work is needed that focuses on the study of interfacial strain, surface effects, and the generation of chemical pressure through doping, to be able to stabilize the skyrmion phase in B20 thin films, while simultaneously increasing the magnetic transition temperature.

Finally, we would like to stress that the analysis of polarization-dependent EXAFS performed here to investigate the local strain at the MnSi/Si(111) interface can be widely employed, e.g., to study superconducting lattices\cite{FogelPRL2001,TangNatPhys2014} or heterostructures incorporating topological insulators.\cite{VobornikNanoLett2014,LiPRL2015}
This methodology provides invaluable information about the local interfacial structure, and may allow to explore its correlation with their fascinating electronic and magnetic properties.


\acknowledgments 
We acknowledge Diamond Light Source for beamtime on B18
under proposal SP-10243, and financial support by the John Fell
Oxford University Press (OUP) Research Fund. TH and SLZ acknowledge
support by the Semiconductor Research Corporation (SRC), and AAB
partial financial support by Diamond, EPSRC and Wadham College, Oxford. 
RC acknowledges the Israel PBC fellowship for support. AK acknowledges the support of the Israel Science Foundation (grant 1321/13).
CP acknowledges financial support through DFG TRR80 and ERC AdG (291079, TOPFIT).
We thank N.-J.\ Steinke and L. B.\ Duffy for help with the XRD/XRR
measurements, and R.\ Boada and P.\ B\"oni for fruitful discussions.


%

\end{document}